\documentclass[11pt]{article}%

\usepackage[english]{babel}
\RequirePackage[T1]{fontenc}

\usepackage{comment}
\usepackage{color}
\usepackage{parskip}
\usepackage{appendix}
\usepackage{amssymb}
\usepackage{booktabs}
\usepackage{comment}
\usepackage{graphicx}
\usepackage{caption}
\usepackage{subcaption}
\usepackage{mathrsfs}
\usepackage{tabularx}
\usepackage{arydshln}
\usepackage{amsmath}
\usepackage{multirow}
\usepackage{multicol}
\usepackage{adjustbox}
\usepackage{authblk}
\usepackage{relsize}
\usepackage{wrapfig}
\RequirePackage{mathptmx}      
\RequirePackage{flushend}
\RequirePackage[numbers,sort&compress]{natbib}
\RequirePackage[colorlinks,citecolor=blue,urlcolor=blue,linkcolor=blue]{hyperref}

\usepackage[utf8]{inputenc}
\usepackage[T1]{fontenc}
\usepackage{amsmath}
\usepackage{graphicx}
\usepackage[scale=0.8]{geometry}
\usepackage{lineno}
\usepackage{xcolor}
\def\babar{\mbox{\slshape B\kern-0.1em{\smaller A}\kern-0.1em B\kern-0.1em{\smaller A\kern-0.2em R~}}}

\newcommand{\FCCee}{$\mathrm{FCC_{ee}}\;$}

\usepackage{sectsty}

\sectionfont{\fontsize{13}{14}\selectfont}
\subsectionfont{\fontsize{11}{12}\selectfont}

\makeatletter
\renewcommand\@author{
    \AB@authlist\\[\affilsep]
    \begin{multicols}{2}
        \AB@affillist
    \end{multicols}

    }
\def\@maketitle{
  \begin{center}
  \let \footnote \thanks
    {\Large\bfseries\@title}
    {\normalsize
      \begin{center}%
        \baselineskip=12pt
        \@author
      \end{center}\par}
    {\large \@date}
  \end{center}%
}
\makeatother

\usepackage{amsmath}
\usepackage{comment}
\usepackage{graphicx}

\title{\flushright{\small DPHEP-2025-01} \\ \centering\huge Data Preservation in High Energy Physics \\ \vspace{0.3cm}\normalsize DPHEP Collaboration }
\author[11]{Alexandre~Arbey}

\author[5]{Jamie~Boyd}

\author[19]{Daniel~Britzger}

\author[24]{Concetta~Cartaro}

\author[14]{Gang~Chen}

\author[25]{Gabor~David}

\author[4]{Dmitri~Denisov}

\author[6]{Cristinel~Diaconu}

\author[5]{Dirk~Duellmann}

\author[7]{Marcus~Ebert}

\author[9]{Eckhard~Elsen}

\author[5]{Jacopo~Fanini}

\author[27]{Dillon~S. Fitzgerald}

\author[17]{Benjamin Fuks}

\author[5]{Gerardo~Ganis}

\author[9]{Achim~Geiser}

\author[15]{Takanori~Hara}

\author[26]{Lukas~Heinrich}

\author[28]{Michael~D.~Hildreth}

\author[3]{Julie~M.~Hogan}

\author[1]{Henry~Klest}

\author[18]{Sabine~Kraml}

\author[4]{Eric~Lan\c{c}on}

\author[21]{Clemens~Lange}

\author[10]{Kati~Lassila-Perini}

\author[9]{Sergey~Levonian}

\author[2]{Dietrich~Liko}

\author[13]{Chiara Mariotti}

\author[16]{Zach~Marshall}

\author[28]{Thomas~McCauley}

\author[12]{Fran\c cois~Le~Diberder}

\author[5]{Jean-Yves~Le Meur}

\author[8]{Gerald~Myatt}

\author[4]{Maxim~Potekhin}

\author[7]{Michael~Roney}

\author[5]{Pablo~Saiz}

\author[22]{Heidi~Schellman}

\author[5]{Jose~Benito~Gonzalez~Lopez}

\author[5]{Matthias~Schröder}

\author[5]{Ulrich~Schwickerath}

\author[5]{Tim~Smith}

\author[9]{David~South}

\author[23]{Giordon~Stark}

\author[5]{Tibor~\v{S}imko}

\author[20]{Jan~Timmermans}

\author[19]{Andrii~Verbytskyi}

\author[5]{Arne~Wiebalck}

\author[12]{Zhiqing~Zhang}

\affil[1]{\footnotesize Argonne National Laboratory, USA }
\affil[2]{\footnotesize Austrian Academy of Sciences, Austria}
\affil[3]{\footnotesize Bethel University, USA}
\affil[4]{\footnotesize Brookhaven National Laboratory, USA }
\affil[5]{\footnotesize CERN, Geneva, Switzerland}
\affil[6]{\footnotesize CPPM, Aix Marseille Universit\'e, CNRS/IN2P3, France}
\affil[7]{\footnotesize Department of Physics and Astronomy, Univ. of Victoria, Canada}
\affil[8]{\footnotesize Department of Physics, University of Oxford, United Kingdom}
\affil[9]{\footnotesize Deutsches Elektronen Synchorotron, DESY, Hamburg, Germany}
\affil[10]{\footnotesize Helsinki Institute of Physics, Finland}
\affil[11]{\footnotesize IP2I, Université Claude Bernard Lyon 1, CNRS/IN2P3, France}
\affil[12]{\footnotesize IJCLab, Paris-Saclay University, CNRS/IN2P3, France}
\affil[13]{\footnotesize INFN Torino, Italy}
\affil[14]{\footnotesize Institute of High Energy Physics, China}
\affil[15]{\footnotesize KEK IPNS / SOKENDAI, Japan} 
\affil[16]{\footnotesize Lawrence Berkeley National Laboratory, USA}
\affil[17]{\footnotesize LPTHE, CNRS, Sorbonne Université, France}
\affil[18]{\footnotesize LPSC, Universit\'e Grenoble-Alpes, CNRS/IN2P3, France}
\affil[19]{\footnotesize Max Planck Institute for Physics, M\"unich, Germany}
\affil[20]{\footnotesize Nikhef, Amsterdam, Netherland}
\affil[21]{\footnotesize Paul Scherrer Institute, Switzerland}
\affil[22]{\footnotesize Physics Department, Oregon State University, USA}
\affil[23]{\footnotesize Santa Cruz Inst. for Particle Physics, Univ. of California,  USA}
\affil[24]{\footnotesize SLAC National Accelerator Laboratory, California, USA}
\affil[25]{\footnotesize Stony Brook University, USA}
\affil[26]{\footnotesize Technical University Munich, Germany}
\affil[27]{\footnotesize University of Michigan, USA}
\affil[28]{\footnotesize University of Notre Dame, Indiana, USA}

\begin{document}
\maketitle

\begin{abstract}
Data preservation significantly increases the scientific output of high-energy physics experiments during and after data acquisition. For new and ongoing experiments, the careful consideration of long-term data preservation in the experimental design contributes to improving computational efficiency and strengthening the scientific activity in HEP through Open Science methodologies. This contribution is based on 15 years of experience of the DPHEP collaboration in the field of data preservation and focuses on aspects relevant for the strategic programming of particle physics in Europe: the preparation of future programs using data sets preserved from previous similar experiments (e.g.\ HERA for EIC), and the use of LHC data long after the end of the data taking. The lessons learned from past collider experiments and recent developments open the way to a number of recommendations for the full exploitation of the investments made in large HEP experiments.\footnote{This document has been submitted to the European Strategy for Particle Physics 2026. }
\end{abstract}

\section{Introduction and context }

The need for data preservation demands long-term collaboration to support the continuing analysis program of complex data produced by large experiments. This was highlighted at the conclusion of operation of several large collider experiments at the beginning of this century, in the years before the LHC began.  A study group was convened and produced in 2009 a report~\cite{DPHEPStudyGroup:2009gfj} that proposed a coordinated initiative in HEP, further detailed in a 2012 blueprint~\cite{DPHEPStudyGroup:2012dsv}. The document outlined initial data preservation projects and solidified the collaborative framework. 

The DPHEP collaboration, launched in 2013 by about ten national funding agencies and CERN, partnered with ICFA to facilitate discussion, consensus, and knowledge sharing in HEP data preservation and governance. It is a unique forum to discuss R\&D for shared preservation tools, align projects, and develop sustainable, cost-effective long-term strategies. Additionally, DPHEP promotes data preservation within the HEP community and among key stakeholders. DPHEP issues global reports~\cite{DPHEP:2015npg,DPHEP:2023blx} and holds regular collaboration workshops. 

It should however be noted that, beyond the success stories described below and surveyed by DPHEP, the overall observation is that \textit{the heritage of complex high-energy physics experiments is in general lost\footnote{It is estimated that only a few percent of the experiments in HEP are connected and active on data preservation issues. One can base this estimate on the CERN experiments grey-book when compared to active experiments involved in DPHEP.}}. Beyond the missed opportunities and given the large investments behind these data sets, this situation does not shed the best light on the place of research in society and may suggest that HEP scientists do not recognize or fully appreciate the value of preserved data as a key research product.

\section{Data preservation concepts}
Across the scientific data landscape, HEP data acquisition and processing presents several outstanding features: size, complexity, distributed and experiment-optimised computing models. It is structured by ``events'', that are time stamped and represent the smallest unit of processing. One of the central computing optimization goals is to extract as much information as possible from that provided by the detector within an event. To make it meaningful, the information needs to be reconstructed and interpreted. Those enhanced events are selected, then grouped in statistical ensembles (histograms), and compared to the simulation and theory. 
Finally, the systematic uncertainties are estimated and the physics measurements are extracted and summarized in publications. 
Although the full information is needed to qualify and reproduce the results, only a small fraction of the data sample constitutes the final analysis.
Collider experiments have broad physics programs, and the experimental collaborations can only publish a fraction of the possible physics analyses during their lifetimes.
This leaves a lot of room to explore and it is not uncommon for HEP experiments that analyses appear that go beyond initial physics analysis plans. 

It is important to note that the term {\it data}, is not restricted to files on disk, but comprises all digitally encoded
~
information which has been created as a result of the planning, running and exploiting of an experiment. It includes therefore software at all levels (analysis scripts as well as the minimally requested reprocessing programs), documentation in all forms, indexing and meta-data, calibration databases etc..  Similarly, {\it preservation} is not restricted to securely storing  data, but comprises all it takes for keeping the data alive, accessible and usable, across the ever changing IT infrastructure and technologies, including data and software quality procedures, computing frameworks etc.  Therefore, the term ``Data Preservation'' is inclusive and covers all necessary ingredients to ensure extraction of original results from experimental data well beyond the period of data taking.

\section{Data Preservation benefits}
The benefits of Data Preservation (DP) were predicted in the initial DPHEP document~\cite{DPHEPStudyGroup:2009gfj}. Long-term continuation and extension of scientific programs maximise existing data's potential. The JADE data rescue, leading to a significantly improved measurement
of the strong coupling, is an iconic example\footnote{S.~Bethke and A.~Wagner,
\href{https://doi.org/10.1140/epjh/s13129-022-00047-8}{Eur. Phys. J. H \textbf{47} (2022), 16}
 [\href{https://arxiv.org/abs/2208.11076}{arXiv:2208.11076} [hep-ex]].}.
Reusing past data can lead to advancements through new theories or techniques\footnote{The case for data reinterpretation and its impact on the preservation is addressed in a community paper submitted to ESPP.}. These sophisticated analyses benefit from full statistical power and enable unique cross-collaboration opportunities to reduce uncertainties, refine measurements, and make new discoveries. Preserving data also supports education, training, and outreach, engaging students in high-energy physics research and attracting new talents. With a perspective of 15 years, the benefits of data preservation have been explored in the past few years through workshops and a general report~\cite{DPHEP:2023blx}. A short account is given below.
\begin{figure}[hhh]
\centering
\includegraphics[width=1.05\textwidth,clip]{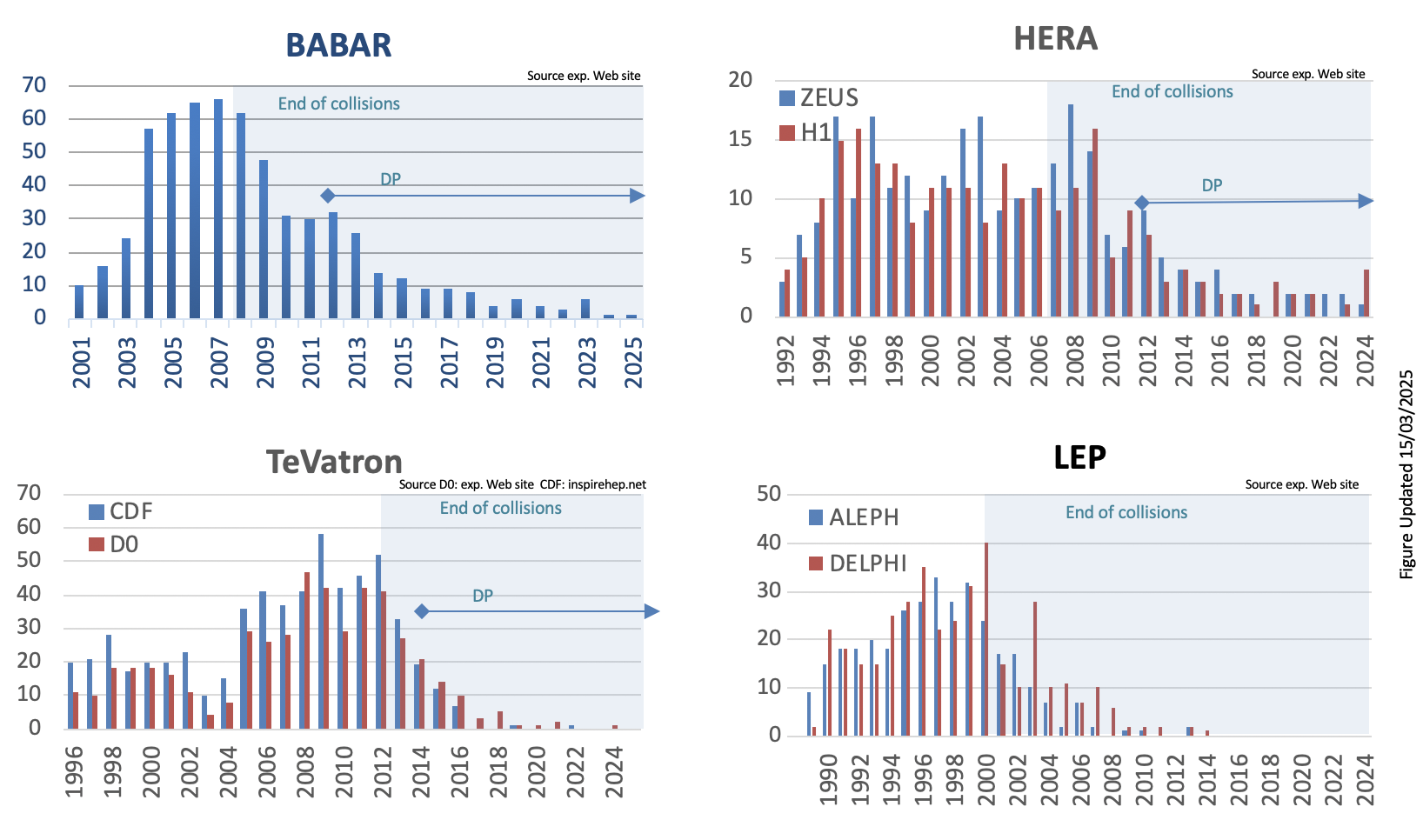}
\caption{Examples of experiments' scientific output. The period after the end of the data taking is indicated by the shaded area. The periods where the collaborations relied on computing projects newly designed for DP is show by an arrow. The period of scientific production is comparable with the data acquisition duration.}
\label{benefits}       
\end{figure}
\subsection{Enhanced scientific outcome }
The main indicator for a successful DP remains the publication record. The experience acquired with large collider experiments demonstrates that the HEP data is exploited in the long term: systematically for at least five years after the end of data taking, and very often for 15 years (e.g.\ large parts of the LEP data are still alive today, 25 years after the end of the data taking). The reasons are multiple: measurements take longer than initially planned, results are improved due to new calculations, require further investigations due to intriguing or discrepant results from other experiments, new ideas of analyses emerge, better methodologies become available\footnote{A striking example is the use of machine learning to reconstruct the Bjorken variable and unfold cross sections at HERA
(V.~Andreev \textit{et al.} [H1],
\href{https://doi.org/10.1103/PhysRevLett.128.132002}{Phys. Rev. Lett. \textbf{128} (2022) no.13, 132002}
[\href{https://arxiv.org/abs/2108.12376}{arXiv:2108.12376} [hep-ex]]).}, etc.

Several historical experiments can assess the costs and benefits of their DP projects with a long term perspective. It is confirmed that the scientific output often continues long after data collection ends, with significant publications occurring post-data taking, as illustrated in Figure~\ref{benefits}. For instance, the \babar collaboration produced more than a third of its publications and HERA experiments increased by 17\% their publication record based on dedicated DP projects implemented a few years after the end of the data taking. The shaded area in Figure~\ref{benefits} indicates that \babar and HERA collaborations performed data analysis after the end of the data taking over a time period longer than their respective data acquisition period, highlighting the importance of structured DP projects and long-term collaborative models. The overall costs, mainly human resources, are minimal compared to total project costs, confirming the 2009 assumptions about DP's high-quality, low-cost contribution to HEP.\footnote{A more details analysis of cost-benefits balance has been attempted in the latest DPHEP global review~\cite{DPHEP:2023blx}.}

Data preservation is a responsible way to manage public investment. 
It enhances education and training for future generations and supports innovative teaching methods. 
The HEP community can use it to prepare for future projects like FCC, EIC, future $b$-factories, and long-term LHC analysis. With HL-LHC data collection ending in 2041, analyses and publications using those data can be expected in the 2050s and possibly beyond. 
Since the HL-LHC would be the last high-energy $pp$-collider for at least 30 years in many scenarios, the LHC and HL-LHC data might not be superseded for many decades. 

\subsection{Preservation in symbiosis with Open Data}
While data preservation (DP) and open data (OD) are often mentioned together, they are distinct and should not be confused.   
Preserved data is not necessarily open or findable. An open approach for the whole DP system (in the sense described above) is only  practical or possible if the computing, curation, and support resources are further increased, while the data preservation activity remains necessary.

\begin{wrapfigure}{r}{0.5\textwidth} 
 \setlength{\belowcaptionskip}{-10pt}.
    \centering
    \includegraphics[width=0.5\textwidth]{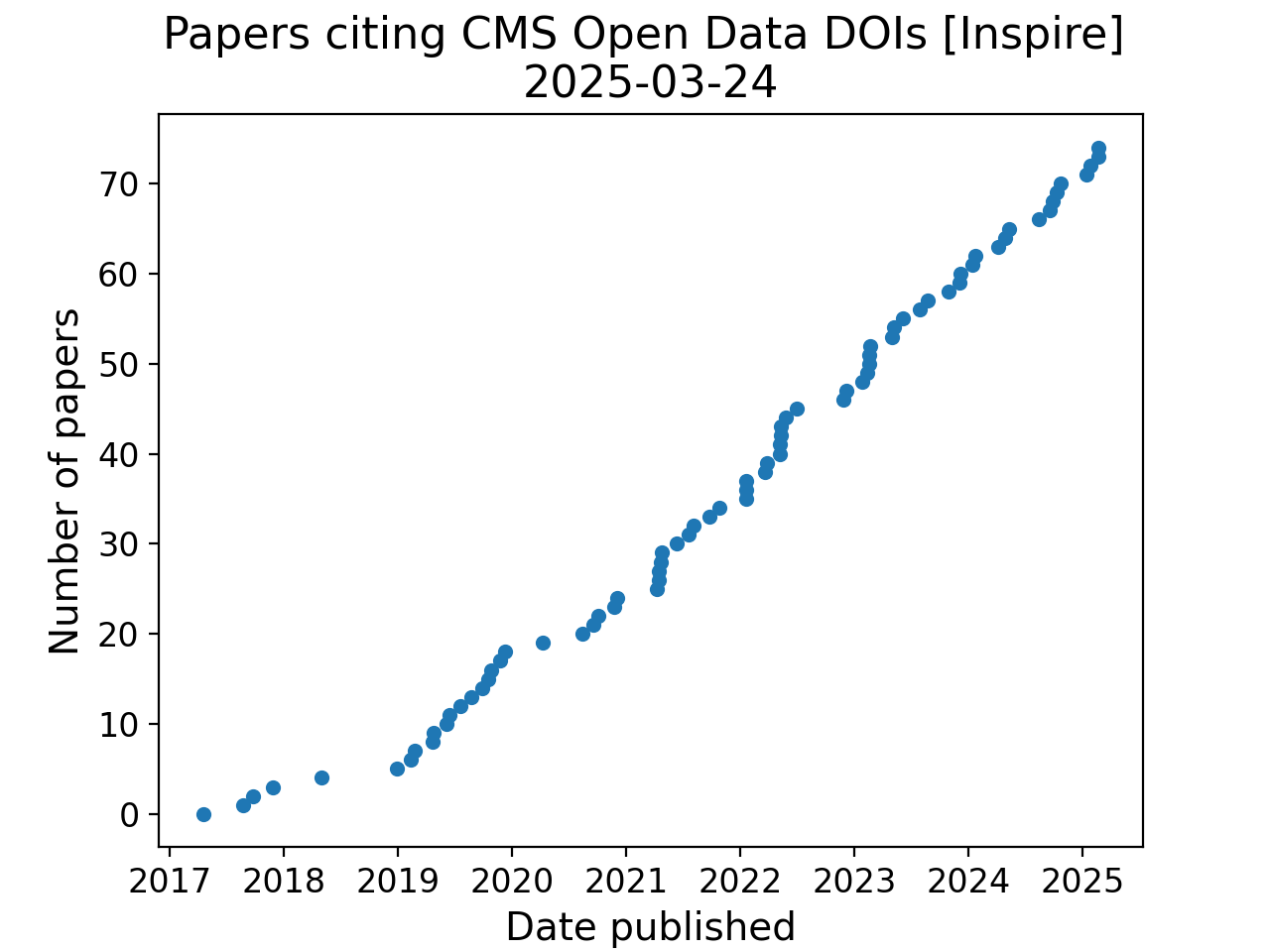}
    \label{npapers}
    \caption{
    Number of publications citing CMS open data records through DOIs, excluding citations by CMS collaboration
and CMS open data contributors (source: \href{https://doi.org/10.5281/zenodo.15078671}{10.5281/zenodo.15078671}).}
\label{npapers}
\end{wrapfigure}
The OD approaches have a clear and solid scientific and technological case: enhancing the scientific output by involving larger communities and simplifying the data analyses methodologies. As an example, Figure~\ref{npapers} shows the scientific output of the CMS open data. Moreover, in the long term, opening access to data is an excellent and probably mandatory way to ensure future usability of valuable experimental data.
Both require clear plans and long term organization and funding, within each collaboration and at international level. While multiple successful examples of DP are already available and mentioned above, the ``DP'' function of ``OD'' approach 
is also increasingly demonstrated by the use of the available data sets.
\footnote{All LHC experiments have open data policies and make notable efforts to release data sets through CERN's \href{https://opendata.cern.ch/}{open data portal}. For instance,  the CMS collaboration released the LHC Run 1 data as part of the Open Data initiative at CERN, thereby ensuring access otherwise not granted by the internal computing environment, while ATLAS has promoted a simplified format, PHYSLITE, for both OD and DP.} 

With the developing OD activity and data availability and the robust Open Science policies adopted at CERN and elsewhere, it is likely that the OD approach constitutes a very promising opportunity to revise, improve and consolidate the DP methodologies for future projects, as has been demonstrated in other fields such astrophysics or gravitational waves. It should however be noted that DP methodologies are indispensable to ensure the back-end for OD in terms of infrastructure, organization and collaboration expertise.

\subsection{Data Preservation technology}
Data preservation was demonstrated to be a test-bed for new and innovative technologies. Indeed, once the ineffectiveness of a ``freezing'' approach (relying on the long-term availability of specific computing hardware) was understood, the dedicated DP projects included techniques such as cloud computing and virtual environments that were not widespread in the HEP context, but have become common in the meantime. 

Most recently, new technologies are being tested in order to address data preservation as a truly complex and heterogeneous system, for which the information is stored  across multiple technologies and facets of data (as defined above). A project proposed by IHEP (Beijing, China) aims to use AI to evaluate the risks of long term failure and identify the critical aspects for data preservation, thereby opening new avenues that may become relevant for running experiments, and potentially leading to versatile and adaptable analysis environments. 

Furthermore, the question of preserving the hardware and hardware-related knowledge is a frontier domain, where common methodologies can be discussed.  The collider experiments in general, and the LHC experiments in particular, incorporate unique, custom made, cutting-edge devices for which fabrication, installation, commissioning and maintenance can be envisaged as aspects to be preserved. Common methodologies are natural for some of those aspects (documentation, CAD software and files, organization forms etc.) while some others can lead to useful cross-fertilization (logistics, safety, maintenance skills, etc.). The common paradigm is the transition from a highly staffed and active period (acquisition and the initial intensive processing and analysis for the data, or construction and commissioning for the hardware) towards a long-term status (data preservation, or long-term maintenance).\footnote{The comparison becomes more obvious if one assumes a high hardware quality and reliability, where technical interventions are significantly reduced after the detector's (and its upgrades') operations are complete: the technical teams move to different activities and their knowledge might decay at a rapid pace, if not preserved in an usable and reproducible form --- in a very similar way as for the data.}

\subsection{Societal impact and cultural heritage }

The HEP data require large investments, but these are spread over many funding agencies, countries and scientists. The vast majority of publications are highly cited and represent unique scientific advances. When the data are preserved for the future, the question of the reproducibility and improvement of published results, as well as the extension of the physics reach is an organizational and technical issue, as described above. However, experimental reproducibility, in the sense of repeating experiments in the same field, is close to zero. The associated data has therefore a significant scientific and historical value and is a key part of the HEP experiments legacy.  

Moreover, reproducing results using the original data (and the associated analysis methodology) can be a very helpful pedagogical exercise. This has been demonstrated for basic examples by the success of the EPPOG masterclasses at high school level using LHC data, but the concept can be extended to include more advanced examples using real data sets from milestone collider experiments. Such extensions are already in active use in university classrooms worldwide. Together with the physics content of the preserved data, the computational and processing methods used for past experiments remain a reference not to be lost, which can be helpful in the design of new experiments. 

When the LHC comes to an end, it might remain the unique hadronic machine at the energy frontier for many years, and its data, together with all the other HEP experiments, will and should be part of the overall heritage of humanity. Moreover, many other large international collaborations now foresee to acquire increasingly large and complex data sets, that need to be preserved in the long term, together with the existing historical data sets. Astrophysics has a long tradition of data preservation in ``virtual observatories'', but still faces new challenges from a significant increase in data volume and complexity. Satellite observations from weather, climate, earth and ocean surveys, as well as medical and social science data sets are highly important. The groups preserving these data can benefit from our experience, and we should try to collaborate and offer tools and knowledge transfer. 

Paradoxically, in an era of intense data-fication of society, scientific data are under threat not only for technical or resource reasons, but more recently also for political reasons. There is a global need to strengthen the critical approach to today's challenges based on scientific evidence. The uniqueness, complexity and innovation encapsulated in the HEP data (in addition to the scientific arguments developed above) justify a common and proportionate conservation action.

\section{Conclusions and recommendations}
Data preservation is an important topic that has proven very effective to enhance the scientific outcome of the HEP experimental programs. It does not emerge spontaneously from the running experiments' computing environments and is not replaced (though largely improved) by the Open Science projects. In order to be successful, it requires dedicated initiatives, technological innovation, adequate resources and solid organization at all levels. 

The following recommendations 
aim to increase the HEP community's awareness of global issues relevant for the full exploitation of the physics potential of HEP data sets. Moreover, these recommendations highlight the assets of European HEP research and its synergy with the global landscape. 
\begin{enumerate}

    \item[] \textbf{Recommendation 1:} Successful data preservation should be promoted and considered as a quality badge of the HEP experiments and host laboratories.  Data preservation is a vector of scientific excellence. Being able to exploit the full potential of an experiment and to use the historical data for education and outreach is a sign of excellence in research and of adequate consideration for the public investment. 
    \item[] \textbf{Recommendation 2:} Strong and coherent action is needed in order to address both data preservation and open science (without assuming that one will solve the ``less important'' problems of the other). Both aspects should be  specifications of the initial design and deliverables of each experiment. The data, defined as the complex ecosystem required to extract scientific results, should be as open as possible and be preserved as long as possible. 
    \item[] \textbf{Recommendation 3:} Dedicated programs enabling the knowledge transfer from preserved data to future experiments need to be encouraged. Concrete examples of tandems of old/future experiments are HERA/EIC and LEP/\FCCee. 
    \item[] \textbf{Recommendation 4:} The investments for data preservation should be better identified and allocated upfront. In particular, the material and human resources, as well as and the organizational impact for a long term LHC data preservation have to be taken into account by the collaborations and CERN.  Data preservation as a laboratory for technology innovation should be taken into account in computing policies. A particular attention should be given to the career aspects, for people involved in data preservation projects. 
    \item[] \textbf{Recommendation 5:} An international organization such as DPHEP is a mandatory ingredient for successful data preservation. It provides reference and support for the long term,  and facilitates the knowledge exchange between participating experiments and projects.  Its action deserves more support, with the objective to preserve as much data as possible and extend data preservation to further experiments. A reinforced effort for trans-disciplinary collaboration can also be envisaged, together with a more prominent role for access rights and intellectual property issues. The operational costs of such an organization are modest, and the long term benefits have been demonstrated. Locating this organization at CERN would be natural, given the world-wide leadership of CERN and its potential to enhance global synergies.
   
\end{enumerate}
\bibliographystyle{utphys} 
{
\bibliography{bib}}

\end{document}